\documentclass[aps,prb,reprint,superscriptaddress]{revtex4-2}
\usepackage{graphicx} 
\usepackage{dcolumn} 
\usepackage{bm} 
\usepackage{hyperref}
\usepackage[mathlines]{lineno}

\usepackage{amsmath}
\usepackage{amssymb}
\usepackage{xcolor}

\begin{document}
	\title{Controllable Majorana vortex states in iron-based superconducting nanowires}
	
	\author{Chuang Li}
	\author{Xun-Jiang Luo}
	\author{Li Chen}
	\affiliation{School of Physics, Huazhong University of Science and Technology, Wuhan, Hubei 430074, China}
	\affiliation{Wuhan National High Magnetic Field Center and Hubei Key Laboratory of Gravitation and Quantum Physics, Wuhan, Hubei 430074, China}
	
	\author{Dong E. Liu}
	\email{dongeliu@mail.tsinghua.edu.cn}
	\affiliation{State Key Laboratory of Low Dimensional Quantum Physics, Department of Physics, Tsinghua University, Beijing, 100084, China}
	
	\author{Fu-Chun Zhang}
	\email{fuchun@ucas.ac.cn}
	\affiliation{Kavli Institute for Theoretical Sciences,  University of Chinese Academy of Sciences, Beijing 100190, China}
	
	\author{Xin Liu}
	\email{phyliuxin@hust.edu.cn}
	\affiliation{School of Physics, Huazhong University of Science and Technology, Wuhan, Hubei 430074, China}
	\affiliation{Wuhan National High Magnetic Field Center and Hubei Key Laboratory of Gravitation and Quantum Physics, Wuhan, Hubei 430074, China}
	
	\date{\today}
	
	\begin{abstract}
		There has been experimental evidence for the Majorana zero modes (MZMs) in solid state systems, which are building blocks for potential topological quantum computing. It is important to design devices, in which MZMs are easy to manipulate and possess a broad topological non-trivial parameter space for fusion and braiding. Here, we propose that the Majorana vortex states in iron-based superconducting nanowires fulfill these desirable conditions. This system has a radius-induced topological phase transition, giving a lower limit to the radius of the nanowire. In the topological phase, there is only one pair of MZMs in the nanowire over a wide range of radius, chemical potential, and external magnetic field. The wavefunction of the MZM has a sizable distribution at the side edge of the nanowire. This property enables one to control the interaction of the MZMs in neighboring vortex nanowires, and paves the way for Majorana fusion and braiding.
	\end{abstract}
	
	\maketitle
	
	\section{Introduction}
	Majorana zero modes (MZMs) have attracted many theoretical and experimental interests given their non-Abelian statistics and their great potential to achieve quantum computation \cite{Kitaev-2003-AnnPhys,Freedman-1998-NAS,DasSarma-2008-RMP}.
	The potential physical realizations of MZM can be roughly classified into two types: 1. end MZMs in one dimension (1D) \cite{Kitaev-2001-PhysUFN,SatoPRL09,DasSarma-2010-PRL,Oreg-2010-PRL}; 2. vortex MZMs in 2D \cite{Read-2000-PRB,FLiang-2008-PRL,SLTD-2010-PRL}.  In early studies, superconducting proximity effect plays an important role for developing the experimentally realizable platforms, for instance, superconductor/semiconductor (SC/Sm) hybrid nanowires \cite{Mourik-2012-Sci,XHongQi-2012-NanoLett,Heiblum-2012-NatPhys,Nichele-2017-PRL,ZHao-2018-NatNano} and Fe chains growing on superconductors \cite{Beenakker-2011-PRB,Yazdani-2014-Sci} for 1D cases and SC/topological-insulator (SC/TI) heterostructures \cite{JJinFeng-2012-Sci,JJinFeng-2016-PRL,hTC-2016-PRB} for 2D cases. However, despite the impressive progress in the epitaxial growth of superconductor \cite{EpitaxyGrowth2015}, the ultra-clean heterogeneous interface requirement still poses various difficulties in fabrication techniques and experimental measurements \cite{Frolov-2020-NatPhys}. Recent studies have revealed that iron-based superconducting materials simultaneously possess superconductivity and topological energy band structure, and hence can support vortex MZMs with no need to fabricate complex heterostructures. This offers a great advantage for the experimental realization and detection of MZMs. For example, the clear zero-bias conductance peak \cite{DHong-2018-Sci,FDongLai-2019-CPL,WHaiHu-2018-NatC,Hanaguri-2019-NatM} and integer quantized Caroli states \cite{FDongLai-2018-PRX,KLingYuan-2019-NatPhys} have been observed in a variety of iron-based superconducting materials \cite{XGang-2016-PRL,DHong-2018-Sci,FDongLai-2018-PRX,ZPeng-2019-NatPhys,DHong-2020-NatCom,FDongLai-2021-PRL}.
	
	On the other hand, it is worth noting that the observation of a zero-bias peak  is a necessary but not sufficient condition for achieving MZMs identification. Perhaps, only the experimental observation of their fusion behaviors and braiding statistics can provide the smoking gun signature. Therefore, the implementation of a Majorana platform optimized for MZMs fusion and braiding is the key to the next milestone. Such an optimized platform should satisfy at least three conditions: 1). a broad parameter range to support well-defined topological degeneracy, 2). an efficient control scheme for MZMs, and 3). a high fidelity and easy readout scheme. For the vortex MZMs, fulfilling the first condition requires precise control of the number of vortex lines in the system, though fine-tuning the parameters is not necessary. The second condition is related to braiding MZMs, which intuitively can be achieved in real space \cite{Ivanov-2001-PRL}; and their physical implementation is very challenging for the current experimental techniques. The promising schemes to perform braiding need to control the neighboring Majorana couplings \cite{Alicea-2011-NatPhys,Sau-2011-PRB,HXiao-2012-EPL,Alicea-2016-PRX,LXin-2016-PRB,Alicea-2017-PRB}. Finally, the fulfillment of the third condition also presupposes the reliable control of Majorana couplings. As far as we know, there are no detailed physical schemes in the vortex MZM platforms to achieve these three conditions.
	

In this work, we propose an iron-based superconducting nanowire setup (Fig.~\ref{fig:NW-VPT}(a)) for fulfilling the above three conditions. Since the repulsive interaction exists among the vortices with finite distances, there is only a single vortex with one pair of MZMs in a wide range of magnetic fields and radius, leading to the unambiguous twofold degenerate ground states. Besides the well-known topological phase transition from varying the chemical potential $\mu$ \cite{Vishwanath-2011-PRL}, we found an additional topological phase transition by tuning the radius $r_0$ of the iron-based superconducting nanowire. Interestingly, unlike the well-known case, the transition occurs at the chemical potential within the bulk gap. This phase transition indicates a lower limit on the radius of the nanowire that can support MZMs. Moreover, there exists a radius range within the topologically nontrivial phase, where the Majorana wavefunction is distributed with a substantial weight near the vortex center as well as on the nanowire lateral surface. The wavefunction distribution in the lateral surface allows a gate-tunable coupling between vortex MZMs through the edge contacts. This radius region can be further extended by introducing a local Zeeman field, for example, on the bottom surface. As this does not affect the top MZM, the bottom MZM does not disappear but moves to the bottom edge. This structure is topologically equivalent to a 2D topological SC possessing a single vortex. It should be noted that an iron-based superconducting nanowire has length $l_0$ and radius $r_0$ in its parameters, which allows for proper separation between the bottom and top MZMs while keeping a large enough gap between the MZMs and the Bogoliubov quasiparticle states to prevent quantum information leakage. In parallel to the benefit of achieving MZMs without requiring proximity effects, our scheme embodies great advantages to meet the requirements of the optimized Majorana platform: first, the geometry of the nanowires will ensure that there are only two MZMs within a certain radius, thus providing well-defined doubly degenerated ground states in the Majorana vortex system; second, the edge MZMs are easily controlled; third, the next step in non-Abelian statistics studies can be carried out with the help of braiding schemes developed from semiconductor nanowires.
	
	The rest of this work is organized as follows: 
	In Sec.~\ref{sec:CriticalRadius}, we discuss the topological phase transition due to the radius of the iron-based superconducting nanowire and the spatial distribution of the Majorana vortex modes. 
	In Sec.~\ref{sec:LocalZeeman}, to obtain tunable MZMs in a wider range of nanowire radius, we induce a local Zeeman field on the bottom surface to push the bottom Majorana vortex to the edge. We investigate the variation of the spectrum and the evolution of the MZM wavefunctions. In Sec.~\ref{sec:EdgeCp}, we study the gate voltage controllable MZMs coupling without and with local Zeeman field. In Sec.~\ref{sec:VortexNumLim}, we discuss the repulsion between vortices in the finite-sized nanowire. This allows the number of MZMs and hence the ground state degeneracy of the nanowires to be regulated by the magnetic field. In Sec.~\ref{sec:Conclusion}, we summarize and conclude our results.
	
	\section{Topological Phase Transition and Majorana vortex wavefunction in iron-based superconducting nanowire} \label{sec:CriticalRadius}
	The Bogoliubov-de$\;$Gennes (BdG) Hamiltonian of the iron-based SC can be written as 
	\begin{align} \label{eq:FeSC-Hmtn}
		H_\text{S}= \begin{pmatrix}
			H_\text{TI} -\mu & \hat{\Delta}_n \\
			\hat{\Delta}^\dagger_n & -H_\text{TI}^* +\mu
		\end{pmatrix}.
	\end{align}
	Here $H_\text{TI}$ is the minimal model describing the topological electronic band structure of Fe(Se,Te) with a band inversion at Z point \cite{XGang-2016-PRL,ZPeng-2019-NatPhys,QShengShan-2019-SciBull}, whose specific form is
	\begin{align}\label{Ham-EL}
		\begin{split}
			H_\text{TI}(\bm{k})&=
			A \hat{\sigma}_x \left( \hat{s}_x\sin{k_x} +\hat{s}_y\sin{k_y} \right) \\
			&+\hat{\sigma}_z \left[ M -B \left( 4-2\cos{k_x}-2\cos{k_y} \right) \right] \\
			& + A_3 \hat{\sigma}_x \hat{s}_z\sin{k_z} -\hat{\sigma}_z B_3\left( 2-2\cos{k_z} \right) \ ,
		\end{split}
	\end{align}
	with $M$, $A $, $A_3$ and $B $, $B_3$ the anisotropic material parameters, $\hat{\boldsymbol{\sigma}}$ and $\hat{\boldsymbol{s}}$ the Pauli matrices acting on the orbits and spin space respectively. In Eq.~\eqref{eq:FeSC-Hmtn}, $\mu$ is the chemical potential measured from the Dirac point of the surface states. The SC term takes the form
	\begin{align}
		\hat{\Delta}_n = -i\hat{s}_y\Delta_0 \left(\tanh(\frac{r}{\xi})e^{i\varphi}\right)^n \ ,
	\end{align}
	where $n$ is 0 (or 1) for the system with no (or one) vortex, $\Delta_0$ is the amplitude of bulk superconducting order parameter, and $\xi= \hbar v_F /(\pi \Delta_0)$ is the superconducting coherence length with the Fermi velocity $v_F$. For the case of $n=1$, it has been well studied that the chemical potential $\mu$ can induce a vortex MZM phase transition with the transition above the bulk band gap \cite{Vishwanath-2011-PRL,QShengShan-2019-SciBull,Lc-2019-SCPMA}. Surprisingly, we found that in the iron-based superconducting nanowire, the size effect, precisely the radius $r_0$, induces an additional topological phase transition.
	
	\begin{figure*}[!th]
		\centering
		\includegraphics[width=17.5cm]{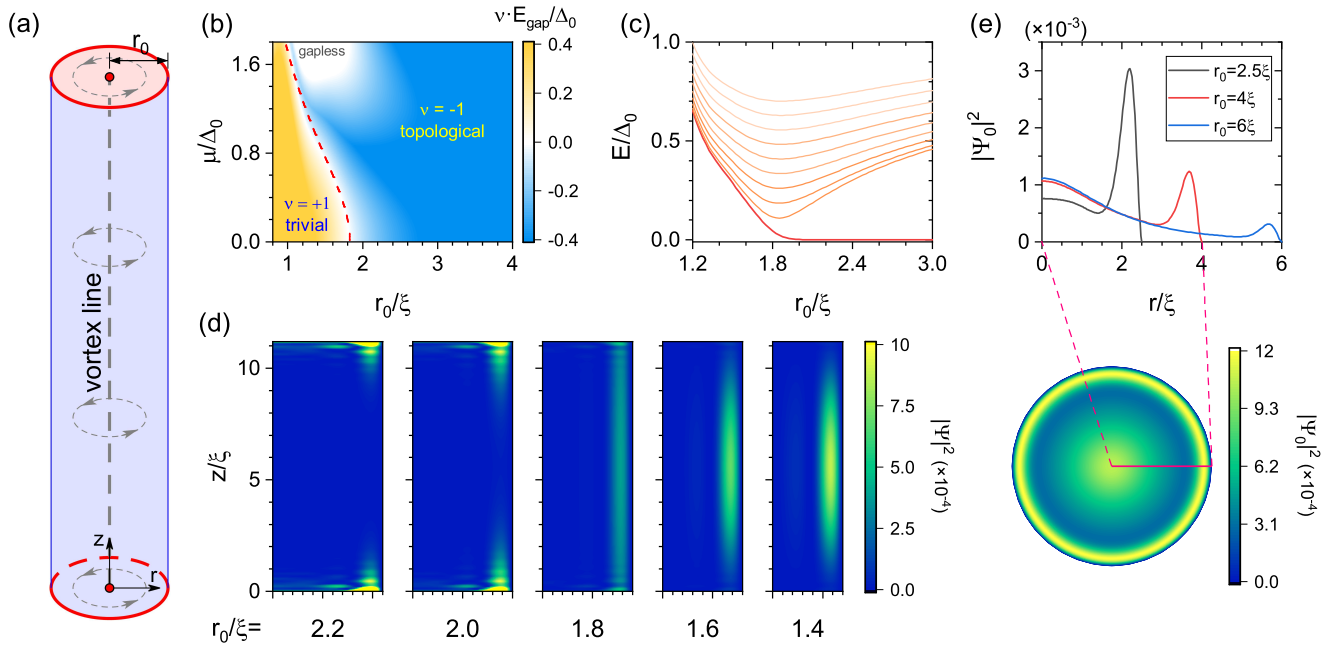}
		\caption{\label{fig:NW-VPT}(color online) The size effect induced topological phase transition in the iron-based superconducting nanowire vortex system described by Eq.~\eqref{eq:FeSC-Hmtn}. (a) The sketch of an iron-based superconducting nanowire model. (b) The product of the topological invariant $\nu$ and the system gap amplitude $E_\text{gap}$ as a function of the nanowire radius $r_0$ and the chemical potential $\mu$. (c) The variation the low-energy spectrum near the topological phase transition at $\mu=0$ with open boundary in z-direction. (d) The probability density $\lvert \Psi \rvert^2(r,z)$ of the lowest energy states in the nanowires with $r_0=2.2\xi,\cdots,1.4\xi$ respectively. (e) The distribution of the MZMs' probability density $\lvert \Psi_0 \rvert^2$ on the circular surface before the phase transition. The parameters used in the calculations are $M=20$~meV, $A=30$~meV$\cdot$nm, $A_3=2.7$~meV$\cdot$nm, $B =-31$~meV$\cdot$nm$^2$, $B_3=4.7$~meV$\cdot$nm$^2$, $\Delta_0=1.8$~meV so that $\xi\approx 5.3$~nm, and 60 layers spaced by 0.6~nm in the $z$-direction.}
	\end{figure*}
	
	To explicitly show this phase transition, we transform the Hamiltonian with one vortex as follows for numerical calculations. Since the band inversion is at $k_z=\pi$, we could adopt a continuous model in the $x$-$y$ plane. Then the system has the continuous rotational symmetry with the total magnetic quantum number $j$, where $j \in \mathbb{Z}$ fulfilling the monodromy of wavefunctions. So we can partition the Hamiltonian into the direct sum of the Hamiltonians with certain $j$ in the $r$-$z$ space as (see Appendix~\ref{sec:App-NumDetail} for details)
	\begin{align} \label{eq:TISC-Hmtn-cyl}
		\mathcal{H}^{(j)}(r,z)= \begin{pmatrix} 
			\mathcal{H}_\text{TI}^{(j)} -\mu & -i\hat{s}_y \Delta_0\tanh(r/\xi) \\
			i\hat{s}_y \Delta_0\tanh(r/\xi) & -{\mathcal{H}_\text{TI}^{(-j)}}^* +\mu
		\end{pmatrix} \ ,
	\end{align}
	where
	\begin{align}
		\begin{split}
			&\mathcal{H}_\text{TI}^{(j)}
			= -iA  \hat{\sigma}_x \left[ \hat{s}_x \partial_r +\frac{i(j+1/2)}{r} \hat{s}_y +\frac{1}{2r} \hat{s}_x \right] \\
			&+\hat{\sigma}_z \left( M +B \left[ \partial_r^2 +\frac{1}{r}\partial_r -\frac{1}{r^2}\left( j-\frac{\hat{s}_z-1}{2} \right)^2 \right] \right) \\ & + A_3 \hat{\sigma}_x \hat{s}_z\sin{k_z} -\hat{\sigma}_z B_3\left( 2-2\cos{k_z} \right) \ .
		\end{split}
	\end{align}
	In the rest of this work, we use the calligraphic font to describes the Hamiltonian for a fixed $j$. The numerical calculations were performed using the Kwant code \cite{Groth-2014-NJP}.
	
	
	A superconducting system with vortex lines along $z$-direction can be considered as a quasi-1D system, which belongs to class D of the Altland-Zirnbauer classification \cite{CCK-2016-RMP}. Note that the particle-hole symmetry gives $\hat{P}\mathcal{H}^{(j)}\hat{P}^{-1}=-\mathcal{H}^{(-j)}$. Therefore, if the system is fully gapped, its topology is characterized by the $\mathbb{Z}_2$ topological invariant $\nu$~ \cite{Kitaev-2001-PhysUFN}
	\begin{align}
		\nu={\rm sgn}\bigg\{\frac{{\rm Pf}[\mathcal{H}_{\rm Mj}^{(0)}(r_0,k_z=0)]}{\text{Pf}[\mathcal{H}_\text{Mj}^{(0)}(r_0,k_z=\pi)]}\bigg\},
	\end{align}
	where $\mathcal{H}^{(0)}_{\rm Mj}$ is the Hamiltonian $\mathcal{H}^{(j=0)}$ written in the Majorana basis. We plot the product of the topological invariant $\nu$ and the system gap amplitude $E_\text{gap}$ as a function of $r_0$ and $\mu$ in Fig.~\ref{fig:NW-VPT}(b), explicitly showing a topological phase transition, characterized by the sign change of $\nu$ and the gap closure. To further understand this phase transition, we consider the case with the open boundary condition along z-direction and a fixed µ = 0, and plot the low energy spectrum (Fig. 1(c)) and the lowest energy wavefunction (Fig. 1(d)) as the nanowire radius decreases. The emergence of the zero-energy mode above the critical radius $r_c\approx 1.8\xi\approx 10$~nm indicates a topological phase transition and the MZM.
	As the radius decreases, the MZM wavefunction shifts toward the edge (also shown in Fig.~\ref{fig:NW-VPT}(e)). When the nanowire radius crosses the critical radius $r_c$, the MZMs on the upper and lower surfaces gradually couple through the lateral surface (Fig.~\ref{fig:NW-VPT}(d)). And these suggest that the phase transition is related to the surface states on the lateral boundary, which can be confirmed by studying the system Hamiltonian at the band inversion point $k_z=\pi$.
	
		\begin{figure}[!th]
		\centering
		\includegraphics[width=8.5cm]{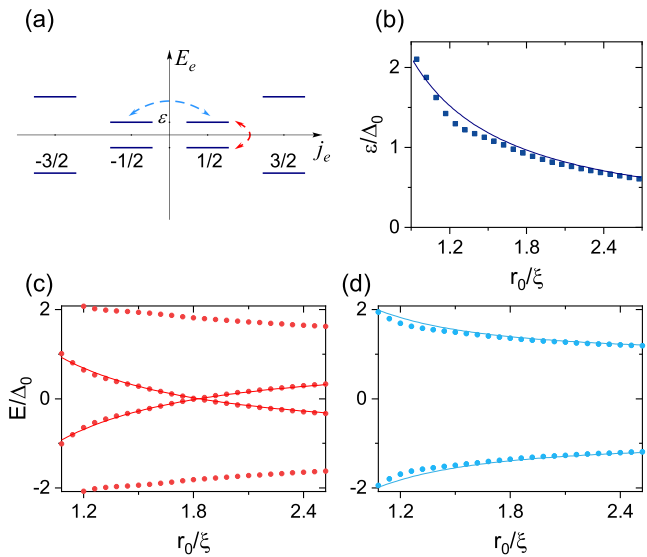}
		\caption{\label{fig:VotxCp-illus}(color online) (a) The spectrum illustration of low-energy states of the electron Hamiltonian part $H_\text{TI}$ from Eq.~\eqref{Ham-EL} with $k_z=\pi$. (b) The lowest positive eigenenergy of the TI for $j_e=1/2$ and $k_z=\pi$. The curve is the approximate energy given by Eq.~\eqref{eq:SideSurf-E0-TI}. (c,d) The low energies (dots) of the superconducting nanowire with $k_z=\pi$ (c) in the presence or (d) in the absence of a single vortex, corresponding to the electron pair indicated by red or cyan dashed line in (a). The curves are approximate analytical energies given by Eq.~\eqref{eq:SideSurf-E0-TSC} and Eq.~\eqref{eq:SideSurf-E0-TSC-NoVortex} respectively. The TI parameters are the same as those in Fig.~\ref{fig:NW-VPT}.}
	\end{figure}
	
	It is noted that the corresponding electron Hamiltonian $H_{\rm TI}$ of Eq.~\eqref{Ham-EL} always respects time-reversal symmetry, and in Eq.~\eqref{eq:FeSC-Hmtn} only the superconducting gap function with one vortex ($n=1$) breaks the time-reversal symmetry. Thus the electronic spectrum, plotted in Fig.~\ref{fig:VotxCp-illus}(a) remains time-reversal invariant, say $E_{j_e}=E_{-j_e}$. In the case of $n=1$, the time-reversal symmetry breaking is reflected in the fact that the Cooper pairs are formed by coupling two electrons with angular momentum $(j_e,-j_e+1)$ (indicated by the red dashed double arrows in Fig.~\ref{fig:VotxCp-illus}(a)). Taking $j_e=1/2$ as an example, the corresponding BdG Hamiltonian projected to $j_e=1/2$ sector states takes the form 
	\begin{align} \label{eq:SideSurf-Hmtn-TSC}
		H_{j_e = \frac{1}{2}} \approx \begin{pmatrix}
			\varepsilon(r_0) \hat{s}_z -\mu \hat{s}_0 & - i\hat{s}_y\Delta(r_0) \\
			i\hat{s}_y\Delta(r_0) & -\varepsilon(r_0) \hat{s}_z +\mu \hat{s}_0
		\end{pmatrix} \ , 
	\end{align} 
	which gives rise to the two lowest eigenenergies
	\begin{align} \label{eq:SideSurf-E0-TSC}
		E_{\pm}(r_0) = \pm \left( \varepsilon(r_0) - \sqrt{\Delta^2(r_0)+\mu^2} \right) \,
	\end{align}
	with
	\begin{align} \label{eq:SideSurf-E0-TI}
		\varepsilon(r_0) = \frac{A }{2r_0}-\frac{B }{2r_0^2}
	\end{align}
	the eigenenergy in the electron spectrum (Fig.~\ref{fig:VotxCp-illus}(b)) (See Appendix~\ref{sec:App-FiniteSize}).
	According to Eq.~\eqref{eq:SideSurf-E0-TSC}, the BdG Hamiltonian of Eq.~\eqref{eq:SideSurf-Hmtn-TSC} closes its gap at $\varepsilon(r_c)=\sqrt{\Delta^2(r_c)+\mu^2}$. In addition to this, the gap closing in other $j_e$ sectors always occurs an even number of times because of particle-hole symmetry. Therefore the topological phase transition is solely determined in the $j_e = 1/2$ sector. We numerically calculate and plot the four eigenenergies closest to zero as a function of $r_0$ in red dots when $\mu=0$ in Fig.~\ref{fig:NW-VPT}(c), which match well with our analytical results (the red solid curves) obtained from Eq.~\eqref{eq:SideSurf-E0-TSC}. 
	Specially, ignoring the spatial variation of $\Delta$, the critical point could be simplified as $r_c \approx A /(2\Delta_0 ) = \pi \xi/2$.
	As a comparison, for the superconducting nanowire without vortices ($n=0$), the electron states will couple their time-reversal partner in SC (indicated by the blue double arrows in Fig.~\ref{fig:VotxCp-illus}(a)), and the energies become
	\begin{align} \label{eq:SideSurf-E0-TSC-NoVortex}
		E_{\pm}(r_0) = \pm \sqrt{\left( \varepsilon(r_0) -\mu\right)^2+\Delta_0^2} \ ,
	\end{align}
	which is always fully opened by the SC gap as shown in Fig.~\ref{fig:VotxCp-illus}(d), corresponding to no topological phase transition and no topological region in the nanowire.
	
	It should be noted that near the phase transition, the lowest energy wavefunctions always distribute at the boundary of the nanowire. This indicates that the related TI surface states of the electron Hamiltonian keep well separated, and therefore the phase transition is not due to the TI surface states coupling. This also means this phase transition only occurs in the superconducting nanowires with topological non-trivial electronic band structures. As the phase transition happens at the lateral surface, we find in Fig.~\ref{fig:NW-VPT}(d) and (e) that when $r_0$ is slightly larger than $r_c$, the MZMs can have a considerable weight on the edge. For example, in Fig.~\ref{fig:NW-VPT}(e) when $r_0$ shrinks to the magnitude of $4\xi$, the distribution of MZMs at the edges is larger than that at the center. This finite distribution of MZMs at edges will give the chance to couple two MZMs from parallel nanowires, which we will discuss later.

	\section{Zeeman field \& Edge MZMs} \label{sec:LocalZeeman}
\begin{figure}[!th]
		\centering
		\includegraphics[width=8.5cm]{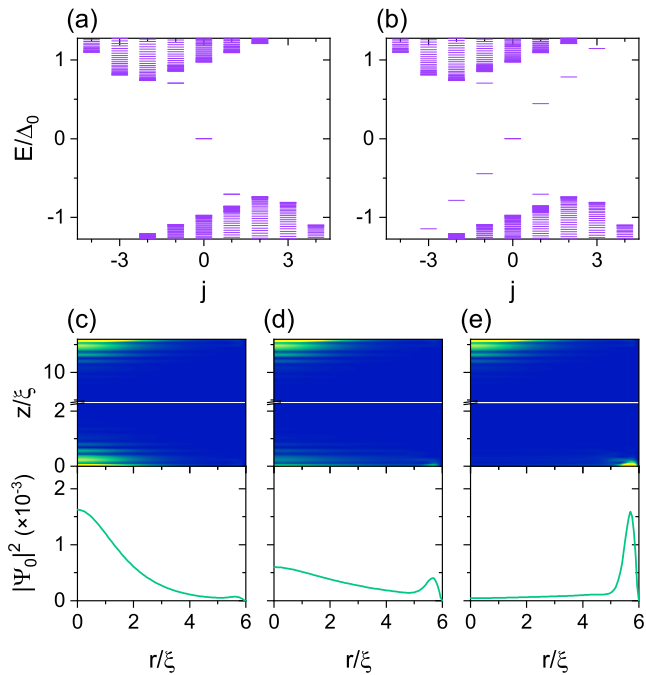}
		\caption{\label{fig:Cyl-WF0-Vz}The spectra of the iron-based superconducting nanowire model Eq.~\eqref{eq:TISC-Hmtn-cyl-Vz} with the bottom layer Zeeman field (a) $V_z=0$ and (b) $V_z=4\Delta_0$ at $\mu= 1.2\Delta_0$. (c$\sim$e) The variation of MZMs' wavefunction $\lvert \Psi_0 \rvert^2$ near the top and the bottom surface corresponding to $V_z=0,\, 2$ and $4\,(\Delta_0)$ in sequence. The lower panels show the probability density $\lvert \Psi_0(r) \rvert^2$ on the bottom surface. Other parameters are the same as those in Fig.~\ref{fig:NW-VPT}.}
	\end{figure}

	In the case of thicker iron-based superconducting nanowires or materials with short SC coherence length, the MZMs wavefunction is mainly concentrated in the vortex center, which is not favorable for manipulation. To improve the controllability of MZMs, we propose to add a local Zeeman field at the bottom surface $\hat{V}_\text{Z}(z) = - V_z \delta(z)\hat{\tau}_z\hat{s}_z$ to push the MZMs to the edge. 
	To avoid bringing large unexpected magnetic fields to the surroundings, this field can be generated by an intralayer ferromagnetic and interlayer anti-ferromagnetic substrate.
	Then the Hamiltonian of the iron-based superconducting nanowire becomes
	\begin{align} \label{eq:TISC-Hmtn-cyl-Vz}
		\mathcal{H}^{(j)}(r,z)= \begin{pmatrix}
			\mathcal{H}_\text{TI}^{(j)} +\hat{V}_\text{Z} -\mu & -i\hat{s}_y \Delta_0\tanh(r/\xi) \\
			i\hat{s}_y \Delta_0\tanh(r/\xi) & -{\mathcal{H}_\text{TI}^{(-j)}}^* -\hat{V}_\text{Z} +\mu
		\end{pmatrix} \ .
	\end{align}
	The localized Zeeman field changes the surface state energy gap from superconducting dominant to magnetic dominant in the bottom surface. In general, when $\lvert V_z \rvert  > \sqrt{\Delta_0^2+\mu^2}$, the superconductivity of this bottom surface will be completely suppressed. In this case, the MZM at the bottom surface will be distributed around the bottom edge, say the boundary between the insulating bottom surface and the superconducting lateral surface. Therefore, the original vortex MZM becomes chiral MZM \cite{FLiang-2008-PRL,PXiaoHong-2019-PRL}. 
	This can be seen from the spectrum and MZMs distribution of the nanowire in Fig.~\ref{fig:Cyl-WF0-Vz}. 
	
\begin{figure}[!th]
		\centering
		\includegraphics[width=3.4in]{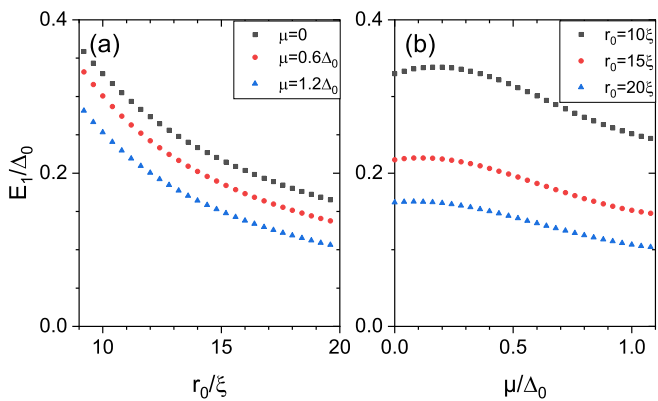}
		\caption{\label{fig:Cyl-E1-Nr}The scatter graphs of the first excited energy $E_1$ of Hamiltonian Eq.~\eqref{eq:TISC-Hmtn-cyl-Vz} with respect to (a) the superconducting nanowire's radius $r_0$ for different chemical potentials, and (b) the chemical potential $\mu$ for different radii of nanowires with $V_z=4\Delta_0$. Other parameters are the same as those in Fig.~\ref{fig:Cyl-WF0-Vz}.}
	\end{figure}

	Without loss of generality, we take a long enough $l_0$ and $r_0=6\xi$ and $\mu=1.2\Delta_0$. Obviously, whether without or with Zeeman filed, the two MZMs always kept degenerated at $j=0$ and $E=0$ as shown in Fig.~\ref{fig:Cyl-WF0-Vz}(a) for $V_z=0$ and \ref{fig:Cyl-WF0-Vz}(b) for $V_z=4\Delta_0$. Meanwhile the bottom and top MZMs are well separated Fig.~\ref{fig:Cyl-WF0-Vz}(c)$\sim$(e) . Tracing the wavefunction of MZMs with increasing $V_z$, the vortex MZM on the lower surface gradually becomes an edge mode (Fig.~\ref{fig:Cyl-WF0-Vz}(c)$\sim$(e)). Meanwhile, further comparing the energy spectrum without and with Zeeman field, we find that the Zeeman field does lead to extra in-gap states (Fig.~\ref{fig:Cyl-WF0-Vz}(b)). Note that the energy difference between the MZMs and the first excited state gives the effective gap, which determines the upper limit of the ambient temperature and the operating speed desired to manipulate the MZMs. Therefore we plot the first excited state energy $E_1$, indicating the effective gap, in Fig.~\ref{fig:Cyl-E1-Nr}(a) versus the radius of the cylindrical model $r_0$ with a fixed wire length. 
	The excited energies become significantly quantized and $E_1$ grows close to half of $\Delta_0$ as the radius $r_0$ shrinks to below $10\xi$. And a lower chemical potential $\lvert \mu \rvert$ leads to a larger energy gap. We also plotted the energy gap versus $\mu$ for different cylindrical radius $r_0$ in Fig.~\ref{fig:Cyl-E1-Nr}(b). As $\lvert \mu \rvert$ increases away from the Dirac point of the topological surface states, the energy gap indeed reduces for any $r_0$ case. We note that even if the radius changes to $r_0 = 20\xi \approx 100$~nm, the energy of the first excited state still has $0.1 \Delta_0 \approx 0.18$~meV.  
	
	Neglecting the bulk states, we analytically calculate the approximate function of the excited energy (see Appendix~\ref{sec:App-EdgeExEnergy})
	\begin{align} \label{eq:E1-R0-mu}
		\tilde{E}_1 
		\approx \frac{\pi}{\tilde{r}_0 \left( 1+ \tilde{\mu}^2 \right)} \quad (\tilde{r} \gg 1)\ ,
	\end{align}
	with the rescaled $\tilde{E}_1= E_1/\Delta_0$, $\tilde{\mu}= \mu/\Delta_0$, $\tilde{r}_0= r_0/\xi$. 
	This approximate function confirms the changing trend of $E_1$ proportional to the inverse of $r_0$ and $\mu^2$, in our iron-based superconducting nanowire system.
	
	Note that in the two-dimensional vortex system, braided MZMs can be achieved by tuning the coupling of the edge MZMs \cite{HXiao-2012-EPL}. In that case, to suppress the coupling of edge chiral MZMs and vortex center MZMs requires increasing the distance between them, while ensuring a considerable energy gap between chiral MZMs and other edge states requires reducing the system size. In two-dimensional topological SCs, these two contradictory conditions are difficult to reconcile, since there is only one adjustable size parameter, the system radius. However, for iron-based superconducting nanowires, these two conditions correspond to two independently tunable parameters, i.e., nanowire length $l_0$ and radius $r_0$, respectively, and thus can be satisfied simultaneously. These reflect the unique advantages of the iron-based superconducting nanowire system.

	\section{Coupled two edge MZMs}\label{sec:EdgeCp}
\begin{figure}[!th]
		\centering
		\includegraphics[width=3.4in]{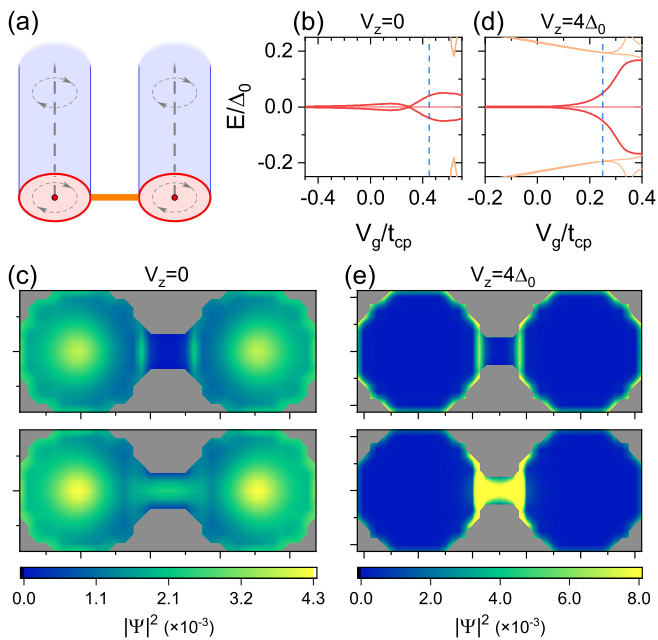}
		\caption{\label{fig:2Cyl-Cp}(color online) (a) Sketch of coupled two MZMs at the bottoms of two parallel nanowires described by Eq.~\eqref{eq:Cp-Hmtn-tot}. For nanowires of small radii $r_0=4\xi$ without Zeeman field, we plot: (b) Spectra of the two-nanowires system as a function of the potential on the connection $V_g$. (c) The probability density of two decoupled ($V_g=-0.5t_\text{cp}$) and coupled ($V_g=0.45t_\text{cp}$, corresponding to the blue dash line in the spectra) Majorana vortex states at the bottom layer. For thicker nanowires ($r_0=5\xi$ here) with Zeeman fields $V_z=4\Delta_0$ at bottom surfaces, we plot: (d) Spectrum variation of the two-nanowires system. (e) The probability density of two decoupled ($V_g=-0.5t_\text{cp}$) and coupled ($V_g=0.25t_\text{cp}$, the blue dash line in the spectra) edge Majorana states. The nanowires' parameters are the same as those in Fig.~\ref{fig:NW-VPT} except that the lattice constant perpendicular (parallel) to the $z$-direction is $\xi/2$ ($\xi/4$) and $M=4$~meV.}
	\end{figure}

	The key ingredient for braiding edge MZMs is the control of the couplings between different Majorana mode~ \cite{HXiao-2012-EPL}. 
	To verify the feasibility of such scheme in the iron-based superconducting nanowires, we explore the coupling of two MZMs at the end of two wires as shown in Fig.~\ref{fig:2Cyl-Cp}(a), and the system Hamiltonian takes
	\begin{align} \label{eq:Cp-Hmtn-tot}
		H_\text{tot}= H_\text{S} +H_\text{N} +H_\text{NS} \ .
	\end{align}
	Here, $ H_\text{S} $ is the iron-based SC system given in Eq.~\eqref{eq:FeSC-Hmtn}.
	Since the lack of rotational symmetry, we now use a 3D tight-binding model with cubic lattice.
	In $ H_\text{S} $ we adjust some parameters to facilitate the calculation, without changing the topological property (see Appendix~\ref{sec:App-NumDetail}).
	The two-vortices SC order parameter is now set as
	\begin{align}
		\Delta(\mathbf{r}) = \Delta_0
		\tanh\frac{r_1}{\xi} e^{i\varphi_1}
		\tanh\frac{r_2}{\xi} e^{i\varphi_2} \ ,
	\end{align}
	where $r_i, \varphi_i$ $(i\in \{1,2\})$ are the horizontal distances and azimuth angles measured from the vertical vortex lines in the centers of the two nanowires respectively. Any closed loop containing $n$ vortex lines changes $\Delta(\mathbf{r})$ by $2n\pi$ phase.
	A gate-tunable semiconductor lead connects the two nanowires, namely the position of the edge components of the two MZMs. We simulate it as a square lattice 
	\begin{align}
		H_\text{N} = t_\text{cp}\hat{\tau}_z \left( 4-2\cos{k_x}-2\cos{k_y} \right) -V_g \ ,
	\end{align}
	and attach it to the superconducting nanowires with
	\begin{align}
		H_\text{NS} = -t_\text{cp} \sum_\alpha \left( c_{\text{N}\alpha}^\dagger \hat{\tau}_z c_{\text{S}\bar{\alpha}} +\text{H.c.} \right) \ ,
	\end{align}
	where $\bar{\alpha}$ indicates the sites at the edge of the nanowires attaching the ends $\alpha$ of the lead.
	The hopping strength is set $t_\text{cp}= 15$~meV, which is corresponding to the material of effective mass about $0.4$ the rest mass of electrons. 
	And $V_g$ is an adjustable on-site potential controlled by the gate voltage.
	
	For the thin superconducting nanowires with diameters $d_0=8\xi \approx 42$~nm so that the MZMs have finite distributions at the edges as we discussed in Sec.~\ref{sec:CriticalRadius}. 
	Adjusting the potential $V_g$ on the connection lead, the variation of the low-energy spectrum of the entire system is plotted in Fig.~\ref{fig:2Cyl-Cp}(b). 
	For $V_g \ll 0$ where the Fermi level is far away from the energy band bottom in the connection lead, the wavefunctions of two MZMs are disconnected as shown in the upper panel in Fig.~\ref{fig:2Cyl-Cp}(c). And their energies keep zeros degenerated with the two vortex center MZMs on the top surface of nanowires.
	Adjusting the potential to $V_g=0.45t_\text{cp}$, as we can see, the two MZMs open a clear energy gap about $0.04\Delta_0 \approx 0.07$~meV, \emph{i.e.}, leaving the degenerated space of zero energy but still away from the excited states to avoid information leakage. Meanwhile, parts of the two MZMs at the bottom surface penetrate the connection lead and couple as shown in the lower panel Fig.~\ref{fig:2Cyl-Cp}(c). 
	While the MZMs on the top are unaffected in this process.
	
	For the thicker superconducting nanowires, the weight of MZMs at the edges will be lower, as well as the coupling by the connection lead. But we could enhance it by adding a Zeeman field $V_z=4\Delta_0$ at the bottom surface as we discussed in Sec.~\ref{sec:LocalZeeman}.
	Here, we take the nanowires of diameters $d_0=10\xi$ with local Zeeman fields as an example.
	The variation of the spectrum and MZMs' wavefunctions are shown in Fig.~\ref{fig:2Cyl-Cp}(d$\sim$e).
	Similarly, the two edge MZMs can be isolated or connected under the adjustment of the gate $V_g$. Though some in-gap interferential edge states are induced by the Zeeman field, under the excited states, the energies of the coupled edge MZMs grow exponentially away from the degenerated zero-energy space as $V_g$ increases.

	%

	\section{Restrictions on the number of vortices}\label{sec:VortexNumLim}
	In the above, we assumed that each superconducting nanowire contains only one vortex. If there are two vortices with the same chirality in the nanowire, the nanowire diameter will limit the distance separating them, which will cause a ﬁnite repulsive potential and increase the free energy of the system. Therefore, it is not surprising that repulsive interactions limit the number of vortices that penetrate the nanowire. The contribution of the vortices to the free energy, through their induced magnetic field $\bm{h}(\bm{r})$, takes the form \cite{Tinkham-2004-book}
	\begin{align}
		\Delta F = \frac{1}{8\pi} \int \left( \lvert\bm{h}(\bm{r})\rvert^2 +\lvert\bm{\nabla} \times \bm{h}(\bm{r})\rvert^2 \right) d^3\bm{r} \ .
	\end{align}
	When there is one vortex in the system with rotational symmetry along $z$-axis, the increased free energy is
	\begin{align} \label{eq:FreeE-1Votx}
		\Delta F_1 = \frac{\Phi_0 L}{8\pi} h_1(\xi) \ ,
	\end{align} 
	where in the range $r \in (\xi,\lambda)$, the induced magnetic field takes the form
	\begin{align}\label{m-vortex}
		h_1(r) \approx \frac{\Phi_0}{2\pi \lambda^2} \left( \ln\frac{\lambda}{r} +0.12 \right) \ ,
	\end{align}
	with $\lambda$ the penetration depth of magnetic field.
	When there are two vortices in the system, the induced magnetic field can be considered as the superposition of the magnetic field induced by each vortex \cite{Tinkham-2004-book}. Therefore, the increased free energy of the two vortices can be estimated to be
	\begin{align} \label{eq:FreeE-2Votx}
		\Delta F_2 = 2 \Delta F_1 + \frac{\Phi_0 L}{4\pi} h_1(2r_0) \ ,
	\end{align}
	where the second term on the right-hand side is from the interaction of the two vortices separating by the maximum distance $2r_0$ inside the nanowire cross-section. 
	
\begin{figure}[!th]
		\centering
		\includegraphics[width=3.4in]{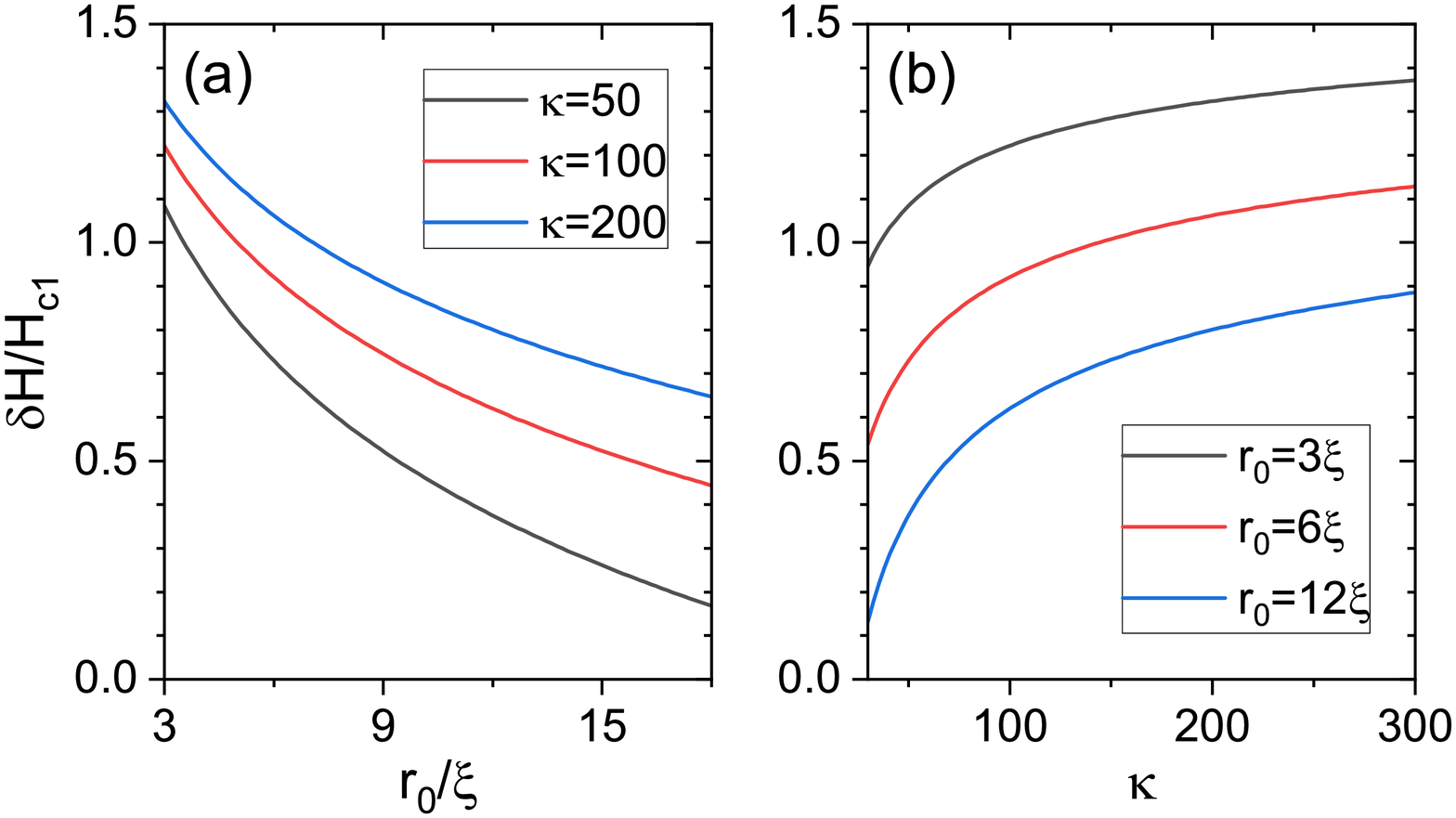}
		\caption{\label{fig:Cyl-dH}The estimated range of magnetic fields $\delta H$ when only a single vortex penetrates the system as a function of (a) the nanowire radius $r_0$ and (b) the dimensionless Ginzburg-Landau parameter $\kappa$.}
	\end{figure}

	On the other hand, when the second vortex starts to appear in the system, the lower critical fields $H_{\rm c1}^{(n=1,2)}$ satisfy the conditions \cite{Tinkham-2004-book} (see Appendix~\ref{sec:App-VortexNumLim})
	\begin{align}
		\Delta F_{1}= \frac{\int \bm{H}_{\rm c1}^{(n=1)} \cdot \bm{h}d^3 \bm{r}}{4\pi}&=H_{\rm c1}^{(n=1)}\frac{\Phi_0 L}{4\pi} \label{eq:free-cond1} \ ,
	\end{align}
	\begin{align}
		\Delta F_2 -\Delta F_{1}&=H_{\rm c1}^{(n=2)}\frac{\Phi_0 L}{4\pi} \label{eq:free-cond2} \ .
	\end{align}
    Substituting Eq.~\eqref{eq:FreeE-1Votx} and \eqref{eq:FreeE-2Votx} to Eq.~\eqref{eq:free-cond1} and \eqref{eq:free-cond2}, we have
	\begin{align}
		\delta H = H^{(n=2)}_{\rm c1} - H^{(n=1)}_{\rm c1} = H^{(n=1)}_{\rm c1} \frac{2h_1(2r_0)}{h_1(\xi)} \ .
	\end{align}
	As we can see, $\delta H$ is not negligible for small size and becomes larger as the distance between vortices becomes more restricted.
	For nanowires with $2r_0 \in (\xi, \lambda)$, according to Eq.~\eqref{m-vortex}, $\delta H$ can be estimated as
	\begin{align}
	    \frac{\delta H}{H^{(n=1)}_{\rm c1}} \approx 2\left[ 1- \frac{\ln 2\tilde{r}_\text{0}}{\ln \kappa} \right]
	\end{align}
	with $\tilde{r}_0 = r_0/\xi$ and $\kappa= \lambda/\xi$ the dimensionless Ginzburg-Landau parameter.
	As shown in Fig.~\ref{fig:Cyl-dH}, materials with larger $\kappa$ and made into thinner nanowires have larger $\delta H$.
	Specifically for Fe(Se,Te), which satisfies $\kappa \approx 10^2$ \cite{Prozorov-2010-PRB} with $r_0$ about dozens of $\xi$, $\delta H$ are of the similar magnitude of $H_{\rm c1}$. So we can control the external magnetic field in the range of $(H_{\rm c1},H_{\rm c1}+\delta H)$ to manufacture the single vortex nanowires and prepare stable edge MZMs.
	
	\section{Conclusion} \label{sec:Conclusion}
	In this work, we propose iron-based superconducting nanowires as a promising platform for achieving controlled MZM. The finite radius of the nanowire limits the number of vortices penetrating it and stabilizes the ground state degeneracy of the Majorana system within a certain range of external magnetic fields. We find that there is a size effect induced topological phase transition when the diameter of the Fe-based superconducting nanowire is reduced to about $\pi \xi$, about 20 nm for Fe(Se,Te) nanowire, which gives a lower limit on the nanowire diameter. When the diameter of the nanowires is about $(\pi \xi, 4\pi\xi)$, the MZMs have a limited distribution not only in the vortex core but also at the edges outside the vortex.  For thicker nanowires, edge MZMs can be obtained by inducing local Zeeman fields, while the excitation energy of the disturbed edge states can be quantified by reducing the radius or decreasing the chemical potential. 
	Edge MZMs in these nanowires can be connected in parallel by tunable semiconductor wires as a key step to achieve MZMs braiding. As of now, with the development of iron-based material growth technology, iron-based superconducting nanowires have been fabricated with a radius of tens of nanometers, which is several coherence lengths \cite{Manashi-2013-ACSNano,WMawKuen-2014-MRE,WXingCai-2018-JAC}. Thus, stable, distinguishable, adiabatically controllable MZMs can be realized in iron-based superconducting nanowires and serve as a cornerstone to study their non-abelian properties.

	\section*{Acknowledge}
	We would like to thank Ching-Kai Chiu, Gang Xu, Yi Zhou, Xiao Hu, Ling-Yuan Kong and Hong Ding for fruitful discussions. X. Liu acknowledges the support of NSFC (Grant No.12074133), NSFC (Grant No.11674114) and National Key R\&D Program of China (Grant No. 2016YFA0401003). F.-C. Zhang is partially supported by NSFC grant No. 11674278, and by the Priority Program of Chinese Academy of Sciences, grant No. XDB28000000. D. E. Liu is supported by NSF-China grant No. 11974198.
	\appendix
	\section{Numerically Solving the Fe(Se,Te) Model Hamiltonian} \label{sec:App-NumDetail}
\begin{figure}[!th]
		\centering
		\includegraphics[width=2.3in]{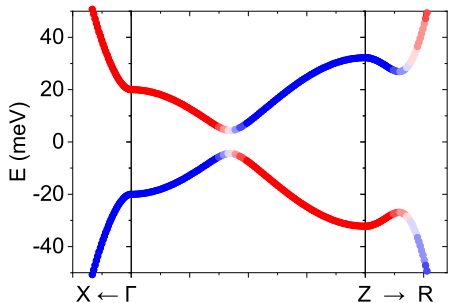}
		\caption{\label{fig:TI-E-k}The band structure of the TI part in the iron-based SC described by Eq.~\eqref{eq:TI-Hmtn-k}. The color maps orbital index $\langle \hat{\sigma}_z \rangle$, showing a band inversion at Z point.}
	\end{figure}

	In Eq.~\eqref{eq:TISC-Hmtn-cyl}, we model the topological surface states in iron-based superconducting nanowires using a TI Hamiltonian which can be written  as \cite{LChaoXing-2010-PRB}
	\begin{align} \label{eq:TI-Hmtn-k}
		\begin{split}
			H_\text{TI}(\mathbf{k})=& \hbar v \hat{\sigma}_x \hat{\boldsymbol{s}} \cdot \mathbf{k}_\parallel +\hat{\sigma}_z(M-B  k_\parallel^2) \\ 
			&+\frac{\hbar v_z}{c} \hat{\sigma}_x\hat{s}_z \sin k_z c -\frac{B_3}{c^2}\hat{\sigma}_z(2-2\cos k_zc) \ ,
		\end{split}
	\end{align} 
	where we adopted long-wavelength approximation in the plane of $ \mathbf{k}_\parallel=(k_x,k_y) $ for subsequent calculations. Here, $c$ is the effective lattice constant along $z$-direction. $\hat{\boldsymbol{\sigma}}$ and $\hat{\boldsymbol{s}}$ are the Pauli matrices acting in the spin and orbital space respectively. 
	$v$, $v_z$, $B $, $B_3$ and $M$ are the material parameters in TI part. 
	The coefficient $v$ ($v_z$) equals the Fermi velocity of the topological surface states perpendicular (parallel) to $z$-direction and $M$ indicates the half of the gap at $\Gamma$ point.
	In order to provide more precise results for experiments, we use the parameters of an anisotropic strong TI model close to the practical band structure in Fe(Te,Se) \cite{ZPeng-2019-NatPhys}, whose bulk dispersion is shown in Fig.~\ref{fig:TI-E-k}, revealing a band inversion occurring at Z point and expected to generate topological surface states surrounding the nanowire's surface. Meanwhile, since $\hbar v_z < \sqrt{2MB_3}$, the dispersion exhibits W-shape in $k_z$ direction, which leads the bulk gap $E_{g,b} \approx 4.4$~meV to be smaller than the gap at high-symmetry point Z $E_{g,\text{Z}} \approx 32.2$~meV.
	For a thick cylindrical model, it could be calculated that the vortex phase transition occurs at $\pm \mu_c = \pm v \sqrt{\lvert E_{g,\text{Z}}/B  \rvert}\approx \pm 30.6$~meV \cite{Vishwanath-2011-PRL}, that is, the MZMs exist in the range from $-\mu_c$ to $\mu_c$, which could be verified by the topological invariant $\nu$. 
	While the radius-induced topological phase transition in thin nanowires has been discussed in Sec.~\ref{sec:CriticalRadius}.
	
	In the rest of this section, we will apply Bessel expansion on the simplified Fe(Se,Te) model to calculate the topological invariant $\nu$, eigen-energies, and eigen-wavefunctions.
	We need to adopt $k$-space for $z$-direction (the direction of vortex line) when solve the topological invariant, while use full real space for eigen-states. Taking the former for example, we rewrite the TI bands part in the Fe(Se,Te) model Eq.~\eqref{eq:TI-Hmtn-k} into real space cylindrical coordinate system as
	\begin{align} \label{eq:TI-Hmtn-rfk}
		\begin{split}
			H_\text{TI}(r,\varphi,k_z)=
			& \hat{\sigma}_z \left[ M+B \left( \partial_r^2 +\frac{1}{r}\partial_r +\frac{1}{r^2}\partial_\varphi^2 \right) \right] \\
			& -i \hbar v \hat{\sigma}_x e^{-i\varphi \hat{s}_z} \left( \hat{s}_x +\frac{1}{r} \hat{s}_y \partial_\varphi \right) +H_z(k_z) 
		\end{split}
	\end{align}
	Inducing the SC term $\hat{\Delta}e^{i\varphi}$, we obtain the entire Hamiltonian Eq.~\eqref{eq:TISC-Hmtn-cyl} describing an iron-based SC vortex system. To simplify the following calculations, we could reduce the angular dimension using the continuous rotational symmetry $[\hat{J}_z, \mathcal{H}_\text{cyl}]=0$ in this cylindrical nanowire, with the $z$-component of total angular momentum $\hat{J}_z= \hat{L}_z +\hbar \hat{\tau}_z(\hat{s}_z-1)/2$ where the orbital part $\hat{L}_z= -i\hbar \partial_\varphi$ and $\hat{\boldsymbol{\tau}}$ are the Pauli matrices in the particle-hole space. The wavefunctions take the form $\hat{U}^{(j)}_\varphi \Psi^{(j)}(r)$ where $\hat{U}^{(j)}_\varphi= \exp\{i[j-\hat{\tau}_z(\hat{s}_z-1)/2]\varphi\}$ with the total magnetic quantum number $j\in \mathbb{Z}$ fulfilling the monodromy of wavefunctions and $\Psi^{(j)}(r)$ is a column vector independent of $\varphi$.
	So we could take the transform $\hat{U}^{(j)}_\varphi$ to reduce $\varphi$ and get
	\begin{align}
		\mathcal{H}_\text{nw}^{(j)}(r,k_z)= \left( \hat{U}^{(j)}_\varphi \right)^{-1} H_\text{nw}(r,\varphi,k_z) \hat{U}^{(j)}_\varphi
	\end{align}
	an effective Hamiltonian of $r$ and $k_z$ corresponding the dark yellow section in Fig.~\ref{fig:NW-VPT}(a) for any certain $j$.
	
	The radial differential operators in $\mathcal{H}_\text{nw}^{(j)}$ could be dealt with by the Bessel expansion method.
	Each component of the wavefunction $\Psi^{(j)}(r)$ with orbital angular quantum number $m$, corresponding the diagonal elements of $[ j-\hat{\tau}_z(\hat{s}_z-1)/2 ]$, can be expanded into a linear combination of a series of $m$-order Bessel functions. And in this Bessel representation, the matrix elements of the corresponding Hamiltonian $ \mathcal{H}_\text{B}^{(j)} $ can be obtained by
	\begin{align}
		\left( \mathcal{H}_\text{B}^{(j)} \right)_{p'q',pq}=
		\langle J^{(m')}_{q'} \lvert \left( \mathcal{H}_\text{nw}^{(j)} \right)_{p'p} \rvert J^{(m)}_q \rangle
	\end{align}
	where $ \rvert J^{(m)}_q \rangle $ is the normalized Bessel function $ J^{(m)}( \alpha^{(m)}_q r/r_0 ) /[r_0 J^{(m+1)}( \alpha^{(m)}_q)/\sqrt{2}] $ with $ \alpha^{(m)}_q $ the $q$-th zero (except the origin) of the $m$-order Bessel function.
	Discarding the high frequency oscillating Bessel functions corresponding large $q$ which has little effect on the low-energy states, then we get the Hamiltonian matrix  $\mathcal{H}_\text{B}^{(j)}$ with finite size.
	
	The topological region of this system can be obtained by regarding the superconducting vortex line as a quasi-1D system with particle-hole symmetry, and the corresponding $\mathbb{Z}_2$ topological invariant can be calculated as \cite{Kitaev-2001-PhysUFN}
	\begin{align}
		\nu = \text{sgn}\{\mathrm{Pf}[\mathcal{H}_\text{Mj}^{(0)}(k_z=0)]\} \text{sgn}\{\mathrm{Pf}[\mathcal{H}_\text{Mj}^{(0)}(k_z=\pi)]\}
	\end{align}
	where $ \mathcal{H}_\text{Mj}^{(j)} $ is the anti-symmetric Hamiltonian matrix under Majorana representation transformed from the $ \mathcal{H}_\text{B}^{(j)} $.
	
	If we keep the terms of $z$-direction in the tight-binding model, the spectrum and the eigen-wavefunctions of this cylindrical iron-based SC system can be calculated by diagonalizing $\mathcal{H}_\text{B}^{(j)}$.
	
	In order to be close to the practical bands in the topological iron-based SC Fe(Te,Se) \cite{ZPeng-2019-NatPhys}, in the calculations of the topological region and eigen-states of the single vortex system, we used an anisotropic TI model with parameter $M=20$~meV, $B =-31$~meV$\cdot$nm$^2$, $B_3=4.7$~meV$\cdot$nm$^2$, $\hbar v= 30$~meV$\cdot$nm, $\hbar v_z= 2.7$~meV$\cdot$nm and $c=0.6$~nm.
	In the SC terms we used $\Delta_0=1.8$~meV, and hence the SC coherence length $\xi \approx 5.3$~nm while the characteristic length of MZMs $\xi_0 \approx 16.7$~nm is used as the unit length. 
	In the 3D tight-binding model when we calculate the coupling of two MZMs, we used a cubic lattice with effective lattice constants $a=\xi/2$ in the horizontal plane and $c=\xi/4$ in $z$-direction. In order to adapt to the reduced energy bandwidth, we adjust the parameter $M=4.0$~meV. The height $l_0=3\xi$ which is enough to isolate two MZMs at opposite ends of the vortex line.
	We used Kwant code \cite{Groth-2014-NJP} to construct the Hamiltonian and the PFAPACK library \cite{Wimmer-2012-TMS} to calculate the Pfaffian.

	\section{Finite Size Effect of TI's lateral surface States} \label{sec:App-FiniteSize}
	In the radius-induced topological phase transition of the iron-based superconducting nanowire, the gap of the TI lateral surface states plays a key role and we will drive it in this section.
	
	Let us force on the TI bands' Hamiltonian Eq.~\eqref{eq:TI-Hmtn-k} at the band inversion point $k_z=\pi/c$
	\begin{align}
		H_\text{TI}
		= \hbar v \hat{\sigma}_x \left( k_x \hat{s}_x +k_y \hat{s}_y \right)
		+(M'-B k^2) \hat{\sigma}_z
	\end{align}
	where $M' = M-4B_3/c^2$ and $M'B >0$.
	In polar coordinate system, we could use the rotational symmetry and rewrite the Hamiltonian as
	\begin{align}
		\begin{split}
			\mathcal{H}_\text{TI}^{(j)}
			=& -i\hbar v \hat{\sigma}_x \left( \hat{s}_x \partial_r +\frac{ij}{r} \hat{s}_y +\frac{1}{2r} \hat{s}_x \right) \\
			&+\hat{\sigma}_z \left( M' +B \left[ \partial_r^2 +\frac{1}{r}\partial_r -\frac{1}{r^2}\left( j-\frac{1}{2}\hat{s}_z \right)^2 \right] \right)
		\end{split}
	\end{align}
	with $j \in \mathbb{Z}+1/2$ now, and divide it into two blocks
	\begin{align}
		\mathcal{H}_\text{TI}^{(j)} &= \mathcal{H}_1^{(j)} \otimes \mathcal{H}_2^{(j)} \\
		\mathcal{H}_2^{(j)} &= -\hat{s}_z \mathcal{H}_1^{(j)} \hat{s}_z
	\end{align}
	by transforming the representation from the basis $\begin{pmatrix}
		c_{a\uparrow} & c_{a\downarrow} & c_{b\uparrow} & c_{b\downarrow}
	\end{pmatrix}^T$ to $\begin{pmatrix} \begin{pmatrix}
			c_{a\uparrow} & c_{b\downarrow}
		\end{pmatrix} & \begin{pmatrix}
			c_{b\uparrow} & c_{a\downarrow} 
	\end{pmatrix} \end{pmatrix}^T$.
	In the $j=1/2$ sector,
	\begin{align}
		\mathcal{H}_1^{(1/2)}
		=& \begin{pmatrix}
			M' +B \left( \partial_r^2 +\frac{1}{r}\partial_r \right) & -i\hbar v \left( \partial_r +\frac{1}{r} \right) \\
			-i\hbar v \partial_r & -M' -B \left( \partial_r^2 +\frac{1}{r}\partial_r -\frac{1}{r^2} \right)
		\end{pmatrix}
	\end{align}
	
	We define the new wavefunction $\bar{\Psi}(r)= \sqrt{r} \Psi(r)$ so that $2\pi \int \lvert \bar{\Psi}(r) \rvert^2 dr= 1$.
	Then, the corresponding Hamiltonian becomes
	\begin{align}
		\bar{\mathcal{H}}_1^{(1/2)}
		=& \begin{pmatrix}
			M' +B \left( \partial_r^2 +\frac{1}{4r^2} \right) & -i\hbar v \left( \partial_r +\frac{1}{2r} \right) \\
			-i\hbar v \left( \partial_r -\frac{1}{2r} \right) & -M' -B \left( \partial_r^2 -\frac{3}{4r^2} \right)
		\end{pmatrix}
	\end{align}
	When the radius $r_0$ is very large, we anticipate there is a zero-energy edge state $\bar{\Psi}_0$ with $\langle 1/r \rangle \rightarrow 0$, and thus the Hamiltonian retains
	\begin{align}
		\bar{\mathcal{H}}_0= \hat{s}_z \left( M' +B \partial_r^2 \right) -i\hbar v \hat{s}_x \partial_r
	\end{align}
	Assuming $\bar{\Psi}_0=  1/\sqrt{2} \begin{pmatrix} \begin{pmatrix}
			-i & \eta
		\end{pmatrix} \begin{pmatrix}
			0 & 0
	\end{pmatrix} \end{pmatrix}^T \phi(r)$ $(\eta= \pm 1)$ which has $\hat{s}_y \oplus \hat{s}_y \bar{\Psi}_0 = \eta \bar{\Psi}_0$, then the Schr$\ddot{\text{o}}$dinger equation gives
	\begin{align}
		\left( M' +B \partial_r^2 +\eta \hbar v \partial_r \right) \phi(r) =0
	\end{align}
	Assuming $\phi(r) \propto e^{\lambda (r-r_0)}$, it becomes
	\begin{align}
		B \lambda^2 +\eta \hbar v \lambda +M' =0
	\end{align}
	When $\eta=-1$, it has two roots with positive real part
	\begin{align}
		\lambda_\pm = \frac{\hbar v \pm \sqrt{(\hbar v)^2-4M'B }}{2B }
	\end{align}
	Hence, $\phi(r) \propto e^{\lambda_+ (r-r_0)} -e^{\lambda_- (r-r_0)}$ is the radius wavefunction of the edge states.
	
	When the $r_0$ is small, the perturbation term
	\begin{align}
		\bar{\mathcal{H}}'= \begin{pmatrix}
			{B /4r^2} & -i{\hbar v/2r} \\
			i{\hbar v/2r} & {3B /4r^2}
		\end{pmatrix}
	\end{align}
	can no longer be ignored.
	For simplicity, we use $\langle 1/r \rangle \approx 1/r_0$ for this exponentially decayed edge state $\bar{\Psi}_0$. Then we have 
	\begin{align}
		\langle \bar{\mathcal{H}}' \rangle = -\frac{\hbar v}{2r_0}+\frac{B }{2r_0^2}
	\end{align}
	Similarly, there is another edge state $\bar{\Psi}_0 \propto \begin{pmatrix}
		\begin{pmatrix}
			0 & 0
		\end{pmatrix} \begin{pmatrix}
			-i & 1
	\end{pmatrix} \end{pmatrix}^T$ with energy $-\langle \bar{\mathcal{H}}' \rangle$ we could obtain from $\mathcal{H}_2^{(j)}$.
	
	Conclusively, as the radius $r_0$ decreases, the zero-energy TI's topological surface states on the side, which are proportional to $\begin{pmatrix}
		1 & 0 & 0 & i
	\end{pmatrix}^T$ and $\begin{pmatrix}
		0 & 1 & -i & 0
	\end{pmatrix}^T$, open a gap $\pm \left( \hbar v/r_0 -B /r_0^2 \right)/2$ in $j= 1/2$ \cite{Governale-2020-NJP}. In the space of these two states and their hole part when the SC vortex is induced, we will find the topological phase transition by the effective Hamiltonian Eq.~\eqref{eq:SideSurf-Hmtn-TSC}.

	\section{Analytic solution of the Zeeman field induced edge MZMs}\label{sec:App-EdgeExEnergy}
	In this section, we are going to solve the MZM localized at the boundary between superconducting topological surface and Zeeman field analytically, and estimate the energy gap between MZMs and excited states by perturbation to show the quantization of excited energies.
	
	Let us focus on the southern hemisphere of the surface of a superconducting nanowire vortex system shown in Fig.~\ref{fig:NW-VPT}(a) and map it to an infinite large 2D disk.
	On the whole disk, there are topological surface states from the TI component, whose effective Hamiltonian is equivalent to a Dirac cone
	\begin{align}
		H_\text{DC}= -i\hbar v e^{-i\varphi\hat{s}_z} \left( \hat{s}_x\partial_r +\frac{1}{r}\hat{s}_y\partial_\varphi \right) \ ,
	\end{align}
	in polar coordinate system. 
	The Zeeman field is mapped into
	$
	\hat{V}'_\text{Z}(r)= -V_z\hat{s}_z\Theta(r_0-r)
	$
	exists in the center of the disk with a radius $r_0$ corresponding to the radius of the original nanowire.
	Outside the Zeeman field is the $s$-wave SC vortex region with the order parameter simplified as 
	$
	\hat{\Delta}'(r) e^{i\varphi}= -i\hat{s}_y \Delta_0 e^{i\varphi} \Theta(r-r_0)
	$
	since it's negligible in the center for a large Zeeman field $V_z \gg \Delta_0$.
	In summary, the Bogoliubov-de$\;$Gennes Hamiltonian of this disk could be written as
	\begin{align}
		H_\text{d}= \begin{pmatrix}
			H_\text{DC}(\mathbf{r})+\hat{V}'_\text{Z}-\mu & \hat{\Delta}'e^{i\varphi} \\
			\hat{\Delta}'^\dagger e^{-i\varphi} & -H_\text{DC}^*(\mathbf{r})-\hat{V}'_\text{Z}+\mu
		\end{pmatrix} \ .
	\end{align}
	
	Again, taking advantage of the rotational symmetry $[\hat{J}_z, H_\text{d}]=0$ and using the method introduced in Appendix~\ref{sec:App-NumDetail}, we could get the Hamiltonian $\mathcal{H}^{(j)}_\text{d}(r,\partial_r)$ for certain total magnetic quantum number $j$. Next we are going to look for the zero-energy wavefunction $\Psi_0(r)= \begin{pmatrix}
		\psi_{e\uparrow} & \psi_{e\downarrow} & \psi_{h\uparrow} & \psi_{h\downarrow}
	\end{pmatrix}^T$ at $j=0$ in the Zeeman field and SC region respectively, which satisfies the Schr$\ddot{\text{o}}$dinger equations
	\begin{align} \label{eq:Pl-SchrEq}
		\mathcal{H}^{(0)}_\text{d}\Psi_0(r)=0 \ .
	\end{align}
	
	Firstly, let us consider the central Zeeman field region $r<r_0$.
	Since the SC term vanishes, the Hamiltonian $\mathcal{H}^{(j)}_\text{d}$ could be divided into two blocks of the electron and the hole parts.
	The general solutions without diverging at the origin $r=0$ are the Bessel functions of imaginary argument
	\begin{align}
		\begin{pmatrix}
			\psi_{e\uparrow} \\ \psi_{e\downarrow}
		\end{pmatrix} \propto \begin{pmatrix}
			I_0(\rho_1 r) \cos\frac{\theta}{2} \\ -iI_1(\rho_1 r) \sin\frac{\theta}{2}
		\end{pmatrix} \ , \
		\begin{pmatrix}
			\psi_{h\uparrow} \\ \psi_{h\downarrow}
		\end{pmatrix} \propto \begin{pmatrix}
			I_0(\rho_1 r) \cos\frac{\theta}{2} \\ iI_{-1}(\rho_1 r) \sin\frac{\theta}{2}
		\end{pmatrix} \ ,
	\end{align}
	with $\theta= 2\arctan[(V_z+\mu)/(V_z-\mu)]$ indicating the spin polarization direction and $\rho_1= \sqrt{V_z^2-\mu^2}/(\hbar v) \approx \sqrt{V_z^2-\mu^2}/(\pi \Delta_0 \xi)$ where $\xi$ is the coherence length of SC.
	For $\rho_1 r \gg 1$, the approximate formula for $m$-order Bessel function of imaginary argument $I_m(\rho_1 r) \approx e^{\rho_1 r}/\sqrt{2\pi \rho_1 r}$ is independent of $m$.
	For $V_z \gg \mu$, $\theta \rightarrow \pi/2$ so that $\cos(\theta/2)=\sin(\theta/2)$.
	Then the zero-energy wavefunction on the Zeeman field side could be written as
	\begin{align} \label{eq:Pl-WF0-Zeeman}
		\Psi_0(r) \propto \frac{1}{\sqrt{2}} \begin{pmatrix}
			C_1 & -iC_1 & C_2 & iC_2
		\end{pmatrix}^T
		\sqrt{\frac{r_0}{r}} e^{\rho_1 (r-r_0)} \ ,
	\end{align}
	for the Zeeman field of sufficient large strength $V_z^2 \gg \Delta_0^2 +\mu^2$ and big size $r_0 \gg \xi_0$. $C_1$, $C_2$ are two undetermined constants.
	
	Secondly, we will concentrate on the SC region where $r>r_0$ without Zeeman field.
	For algebraic simplicity, we first consider the case of $\mu=0$, namely the Fermi level right on the Dirac point. The Hamiltonian in Eq.~\eqref{eq:Pl-SchrEq} could be block anti-diagonalized into two different spin parts, and each of them can be solved separately. For the spin-up component, 
	its solution that converges at infinity is the MZM already proposed in the SC-TI interface model \cite{FLiang-2008-PRL} 
	\begin{align}
		\begin{pmatrix}
			\psi_{e\uparrow} \\ \psi_{h\uparrow}
		\end{pmatrix} \propto \begin{pmatrix}
			1 \\ -i
		\end{pmatrix}
		\phi(r) \ ,
	\end{align}
	with 
	\begin{align}
		\phi(r)=\exp\left[-\frac{\Delta_0}{\hbar v}(r-r_0) \right] \ .
	\end{align}
	While for spin-down components, we could find another zero-energy solution $\begin{pmatrix}
		\psi_{e\downarrow} & \psi_{h\downarrow}
	\end{pmatrix}^T \propto \begin{pmatrix}
		1 & i
	\end{pmatrix}^T r^{-1} \phi(r)$.
	It diverges at $r=0$ but is still valid in this central Zeeman split model.
	Now, matching the wavefunctions at the boundary $r=r_0$ with ones in the central region, we get the coefficients $C_2=-iC_1$ in Eq.~\eqref{eq:Pl-WF0-Zeeman}, and the zero-energy state's wavefunction in SC region for $\mu=0$ case is
	\begin{align}
		\Psi_0(r) \propto \frac{1}{\sqrt{2}} \begin{pmatrix}
			1 & -i\frac{r_0}{r} & -i & \frac{r_0}{r}
		\end{pmatrix}^T \phi(r) \ .
	\end{align}
	
	Then, we could pursuit for the low-energy excited energies with $m \neq 0$ for $\mu=0$.
	The $m \neq 0$ part in Hamiltonian Eq.~\eqref{eq:Pl-SchrEq} can be view as a perturbation
	\begin{align}
		\mathcal{H}'^{(j)}= \frac{j\hbar v}{r} \hat{\tau}_z \hat{s}_y \ ,
	\end{align}
	and the excited energies could be calculated as $E_j= \langle \Psi_0 \lvert H'^{(j)} \rvert \Psi_0 \rangle/\langle \Psi_0 \vert \Psi_0 \rangle$.
	The zero-energy wavefunction has been solved in $\mu=0$, but we could simplify it before the integration.
	Note that, leaving from $r=r_0$, the wavefunction in Zeeman field region is almost exponentially decay with length $\rho_1^{-1} \approx \xi_0 \Delta_0/V_z$, while in SC region the decay length is approximate to $\xi_0$.
	Since $\Delta_0 \ll V_z$, the main part of integration is in the SC side, and we could ignore the Zeeman field part of the wavefunction in calculation.
	Furthermore, when $r_0 \gg \xi_0$, the inverse proportional factor appeared in spin-down components could be neglected.
	Under these approximations, we finally get
	\begin{align}
		\lvert \tilde{E}_j \rvert 
		\approx \frac{j}{\tilde{r}_0+\frac{1}{2}} \ ,
	\end{align}
	with the dimensionless quantities $\tilde{E}_j=E_j/\Delta_0$ and $\tilde{r}_0= r_0/\xi_0= r_0/(\pi \xi)$.
	Therefore, the gap between MZMs and the lowest excited energy $E_1$, which protects the MZMs' information, is approximately inversely proportional to the radius of Zeeman field $r_0$.
	
	For $\mu \neq 0$ in SC region, refer to the wavefunctions we just solved, we could transform the Hamiltonian in Eq.~\eqref{eq:Pl-SchrEq} into the basis $\begin{pmatrix}
		c_\uparrow+ic_\uparrow^\dagger & c_\downarrow-ic_\downarrow^\dagger & c_\uparrow-ic_\uparrow^\dagger & c_\downarrow+ic_\downarrow^\dagger
	\end{pmatrix}^T$, then the solution that converges at infinity could be found via one of the submatrices in the block anti-diagonalized Hamiltonian
	\begin{align}
		\begin{pmatrix}
			\psi_{e\uparrow}+i\psi_{h\uparrow} \\ \psi_{e\downarrow}-i\psi_{h\downarrow}
		\end{pmatrix} \propto \begin{pmatrix}
			C_3J_0(\rho_2 r) +C_4N_0(\rho_2 r) \\
			-C_3J_1(\rho_2 r) -C_4N_1(\rho_2 r)
		\end{pmatrix} \phi(r) \ ,
	\end{align}
	with $\rho_2= \mu/A$ and $J_m$ ($N_m$) is the Bessel (Neumann) function of $n$ order. $C_3$, $C_4$ are two undetermined constants.
	Recalling the boundary conditions $\Psi_0(r_0) \propto \begin{pmatrix}
		1 & -i & -i & 1
	\end{pmatrix}^T$ for $\mu^2 \ll V_z^2-\Delta_0^2$, we could approximate the zero-energy wavefunction as
	\begin{align}
		\Psi_0(r) \propto \frac{1}{\sqrt{2}} \begin{pmatrix}
			\cos\left(\rho_2 (r-r_0) -\frac{\pi}{4} \right) \\
			i\sin\left(\rho_2 (r-r_0) -\frac{\pi}{4} \right) \\
			-i\cos\left(\rho_2 (r-r_0) -\frac{\pi}{4} \right) \\
			-\sin\left(\rho_2 (r-r_0) -\frac{\pi}{4} \right)
		\end{pmatrix} \sqrt{\frac{r_0}{r}} \phi(r) \ .
	\end{align}
	
	Again, we could calculate the excited energies by perturbation theory for $\mu \neq 0$ cases.
	Here, in the integrand we have used an approximation to turn part of the probability density into a standard exponential shape for simplicity of calculation $(r_0/r) \exp[ -2\Delta_0(r-r_0)/(\hbar v) ] \approx \exp[ ( 2\Delta_0/(\hbar v) +r_0^{-1} ) (r-r_0) ]$.
	And finally the results of the excited energies are 
	\begin{align}
		\lvert \tilde{E}_j \rvert
		&\approx \frac{j\left(\tilde{r}_0 +\frac{1}{2} \right)}{\left(\tilde{r}_0 +\frac{1}{2} \right)^2 +\left(\tilde{\mu}\tilde{r}_0 \right)^2} \\
		&\approx \frac{j}{(1+\tilde{\mu}^2) \tilde{r}_0} \ , \label{eq:Cyl-Ana-Em}
	\end{align}
	with the dimensionless quantities $\tilde{E}_j=E_j/\Delta_0$, $\tilde{r}_0= r_0 /\xi_0= r_0/(\pi \xi)$ and $\tilde{\mu}=\mu/\Delta_0$.
	The energy gap $E_1$ is inversely proportional to the Zeeman field's radius $r_0$ while approximately inversely proportional to the square of chemical potential $\mu$. Thus, in order to get a large energy gap to reduce the interaction between MZMs and excited states, it is important to control the chemical potential close to the Dirac point in this system and make the cylinder slenderer.
	
	\section{Vortices in Finite Size Systems}\label{sec:App-VortexNumLim}
	In this section, we are going to investigating the free energy of the superconducting nanowire vortex system, and prove that there is indeed a finite range of the external magnetic field that limits the vortices number to one.
	
	Applying a external magnetic field $H_\text{ext}$ on the superconducting nanowire in the length direction, the first vortex will penetrate the SC when $H_\text{ext}$ is at the nanowire's lower critical field $H_{c1}^{(n=1)}$. And further more, we assume that the second vortex will appear at $H_{c1}^{(n=1)} +\delta H$ (we use $H$ for magnetic field instead of Hamiltonian in this section).
	With the change of $H_\text{ext}$, the Gibbs free energy of the superconducting system is always continuous, and through that, we could relate $H_{c1}$ and $\delta H$ to the vortices' free energies \cite{Tinkham-2004-book}.
	
	In general, if there are $n$ vortices penetrating the nanowire, each with a flux $\Phi$, then the Gibbs free energy $G_n$ can be written as
	\begin{align}
	    G_n= F_n -H_\text{ext}\frac{n \Phi L}{4\pi} \ .
	\end{align}
	Here, $F_n$ is the Helmholtz free energy of the superconductor containing $n$ vortices and $L$ is the length of nanowire.
	The energy of $n$ vortices is $\Delta F_n = F_n -F_0$.
	At $H_\text{ext}=H_{c1}^{(n=1)}$, the continuity of Gibbs free energy requires $G_0=G_1$, and this gives the energy of a single vortex
	\begin{align} \label{eq:VortexEnergy}
	    \Delta F_1 = H_{c1}^{(n=1)} \frac{\Phi L}{4\pi} \ .
	\end{align}
	At $H_\text{ext}=H_{c1}^{(n=1)} +\delta H$, from $G_1=G_2$ we obtain $\Delta F_2- \Delta F_1= (H_{c1}^{(n=1)} +\delta H) \Phi L/(4\pi)$. Here the energy of two vortices consists of $\Delta F_2 = 2\Delta F_1 +F_\text{int}$ with $F_\text{int}$ the interacting term. Then we have 
	\begin{align} \label{eq:VortiesInteraction}
	    F_\text{int}= \delta H \frac{\Phi L}{4\pi} \ .
	\end{align}
	Therefore, there is a finite suitable magnetic field range $\delta H$ as the positive $F_\text{int}$.
	
	Next we will estimate $\delta H$ in a superconducting nanowire with a radius $r_0$ \cite{Tinkham-2004-book}.
	From the Ginzburg-Landau equation and Maxwell equation, we could derive the magnetic field distribution generated by a single vortex $h_1(r)$, which approximates to the $0$-order Hankel function of imaginary argument 
	\begin{align}
		h_1(r) \approx h_0 \left( \ln\frac{\lambda}{r} +0.12 \right) \ ,
	\end{align}
	at $\xi<r<\lambda$ with $\lambda$ the penetration depth of magnetic field and $\xi$ the SC coherence length. $h_0$ is the characteristic magnetic field strength.
	And through this, we could calculate the single vortex energy
	\begin{align}
		\Delta F_1= \frac{\lambda^2 h_0}{4} h_1(\xi) \ ,
	\end{align}
	and the interaction energy between two vortices $F_\text{int}= \lambda^2 h_0 h_1(r_2)/2$, where $h_1(r_2)$ indicates the magnetic field distribution at the second vortex core generated by the first one.
	To minimize the repulsion between two vortices as well as the Gibbs free energy $G_2$, the distance between the two vortices should be maximized, namely $2r_0$ in the cross-section of the nanowire, then
	\begin{align}
		F_\text{int}=\frac{\lambda^2 h_0}{2} h_1(2r_0) \ .
	\end{align}
	
	Substituting them into Eq.~\eqref{eq:VortexEnergy} and Eq.~\eqref{eq:VortiesInteraction}, and according to the approximate shape of field $h_1(r)$, we get
	\begin{align} \label{eq:Cyl-dH}
		\frac{\delta H}{H_{c1}} 
		= \frac{F_\text{int}}{\Delta F_1}
		= \frac{2h_1(2r_0)}{h_1(\xi)} 
		\approx 2\left[ 1- \frac{\ln 2\tilde{r}_\text{0}}{\ln \kappa} \right] \ , 
	\end{align}
	where the diameter, $\tilde{r}_\text{0}= r_0/\xi \in (1/2,\kappa/2)$  and the dimensionless Ginzburg-Landau parameter $\kappa= \lambda/\xi$.
	

%

\end{document}